\documentclass[twocolumn,aps,prb,tightenlines,amsmath,amssymb]{revtex4}
\usepackage{graphicx}
\usepackage{bm}
\usepackage{amssymb}
\usepackage{amsmath}
\usepackage{colordvi} 
\usepackage{color}

\begin{document}

\title{Singlet-triplet relaxation induced by confined phonons in
  nanowire-based quantum dots}

\author{Y.\ Yin}
\email{yin80@ustc.edu.cn.}
\affiliation{Hefei National Laboratory for Physical Sciences at
  Microscale and Department of Physics, 
University of Science and Technology of China, Hefei,
  Anhui, 230026, China}

\date{\today}

\begin{abstract}
  The singlet-triplet relaxation in nanowire-based quantum dots induced by
  confined phonons is investigated theoretically. Due to the
  quasi-one-dimensional nature of the confined phonons, the singlet-triplet
  relaxation rates exhibit multi-peaks as function of magnetic field and the
  relaxation rate between the singlet and the spin up triplet state is found to
  be enhanced at the vicinity of the singlet-triplet anti-crossing. We compare
  the effect of the deformation-potential coupling and the piezoelectric
  coupling and find that the deformation-potential coupling dominates the
  relaxation rates in most cases.
\end{abstract}


\maketitle

\section{Introduction}

As one of the most promising candidates for qubits, spins in semiconductor
quantum dots (QDs) have attracted much attention in the last two
decades.~\cite{loss, elzerman, hanson} Although much progress has been made in
single-electron QDs, an attractive alternative is to base the qubits on
singlet-triplet (ST) states in two-electron QDs.~\cite{petta, petta2, koppen,
  koppen2, barthel, petta3} This is mainly due to two important features of the
ST states. One is the Zeeman-driven ST transition with anti-crossing due to the
spin-orbit coupling (SOC), which enables coherent manipulation of spin
states.~\cite{petta3, johnson, hanson2} The other one is the low ST relaxation
rate which makes the spin polarization remains for a sufficiently long
time.~\cite{fujisawa, sasaki, meunier} As the QDs are either self-assembled ones
sitting on the surface of the substrate or fabricated by confining electrons in
quantum wells by electrodes, bulk phonons in the substrate play an important
role. In low-temperature regime, acoustic bulk phonons in conjunction with the
SOC serve as the main source of the ST relaxation due to the suppression of the
effect of the hyperfine interaction in the presence of large magnetic
field. Previous studies show that at the vicinity of the ST anti-crossing, the
relaxation rate is greatly suppressed due to the mismatch of the phonon emission
wave length to the dot size. Furthermore, the spin-up triplet states which
couple to the singlet ground state through the SOC has a much shorter lifetime
compared to the other two triplet states due to the strong ST mixing between
them.~\cite{shen, wang, golovach, florescu, chaney, climente}

Despite of these progress in two-electron QDs, the complicated fabrication
process makes them difficult to be scaled up, which limits their application
from an industrial point of view. Recently, self-assembled two-electron QDs
embedded in InAs [111] nanowires were fabricated.~\cite{bjork:1058, bjork:1621,
  nilsson:163101} The well-controlled growth process enables parallel production
in massive number with similar properties, which makes these nanowire-based QDs
are more suitable for integration onto silicon and scaling up for large
hierarchical systems. Although similar Zeeman-driven ST transition has been
obtained in nanowire-based two-electron InAs QDs,~\cite{fasth, pfund, pfund2}
the ST relaxation can be quite different due to the properties of the confined
phonons in the nanowires. The nanowires are perpendicular to the substrate,
making the bulk phonons in the substrate are less important than the confined
phonons in nanowires.~\cite{ohlsson:3335, shtrikman:2009} The regular structure
of the nanowires results in the quasi-one-dimensional confined phonons, which
lead to novel properties in optical absorption and transport for the
nanowire-based QDs.~\cite{PhysRevLett.99.087401, prl.104.036801} Therefore,
these confined phonons are expected to have a pronounced impact on the ST
relaxation, which has not been studied to the best of our knowledge. In this
work we will investigate the ST relaxation induced by the confined phonons.

\section{Model and Formalism}

Consider a two-electron elongate QD embedded in InAs [111] cylindrical nanowire
with radius $R$ in the presence of external magnetic field $B$ along the
wire. We model the QD by an anisotropic harmonic potential $V_c(r, z) =
\frac{1}{2} m^{\ast}\omega^2_0 r^2 + \frac{1}{2} m^{\ast} \omega^2_z z^2$ (
corresponding to the effective dot length $d_z=\sqrt{\hbar \pi / m^{\ast}
  \omega_z}$ and dot diameter $d_0=\sqrt{\hbar \pi / m^{\ast} \omega_0}$) with
$z$-axis along the wire. $m^{\ast}$ is the effective mass of the electron. The
single-electron Hamiltonian can be expressed as
\begin{equation}
  H_{\rm{e}} =  \frac{\bm{p}^2}{2 m^{\ast}} + V_c(r, z) + H_{\rm{B}} + H_{\rm{SO}},
\end{equation}
where $H_{\rm{B}} = \frac{1}{2} g \mu_{\rm{B}} \bm{B} \cdot \bm{\sigma}$ is the Zeeman
splitting with $g$, $\mu_{\rm{B}}$ and $\bm{\sigma}$ being the $g$ factor of electron,
Bohr magneton and Pauli matrix respectively. $H_{\rm{SO}}$ represents the SOC
term. In this work, we concentrate on the Rashba coupling which is dominant in
InAs nanowires which has the form $H_{\rm{SO}} = \frac{\gamma}{\hbar} \sigma^y p^z$
with $\gamma$ being the Rashba coupling strength.\cite{fasth, romanon} We assume
the dot diameter $d_0 \ll d_z$ so that only the lowest electron subband in the
radial direction is needed. We also assume the external magnetic field is weak
enough so that its orbital effect can be neglected.

The confined phonons are calculated with isotropic elastic continuum model which
is widely used in the study of nanowires, carbon nanotubes and
nanoparticles.~\cite{cleland, PhysRevLett.71.3577, suzuura, chassaing, yu,%
  komirenko} The nanowire is modeled as an infinite cylinder with the stress
vanishing at the surface of the wire. The displacement field of the ions induced
by the confined phonons can be expressed by the quantized form (in cylindrical
coordinate)
\begin{equation}
  \bm{u}(\bm{r}) = \sum_{\nu q} \sqrt{\frac{\hbar}{2 \rho V \omega_{\nu q}}}
  \bm{\bar{u}}_{\nu q}(r) e^{i l \theta + i q z} ( a_{\nu q} + a^{\dagger}_{\nu
    -q} ),
  \label{eq:uph}
\end{equation}
where $V$ is the volume of the nanowire and $\rho$ is the density. The
eigenmodes $\bm{\bar{u}}_{\nu q}(r)$ and spectrum $\omega_{\nu q}$ are
calculated following Refs.~\onlinecite{auld, stroscio}. Note the due to the
boundary condition, generally speaking, the confined phonon modes have both
longitude and transverse components.

Given the displacement field, both the deformation-potential coupling and the
piezoelectric coupling can be calculated. The deformation-potential coupling is
given by the divergence of the displacement field, which reads
\begin{equation} 
  H^{\rm{D}}_{\rm{ep}}(\bm{r}) = -\Xi \nabla \cdot \bm{u}(\bm{r}),
  \label{eq:def}
\end{equation}
where $\Xi$ is the deformation-potential coupling strength. Note that only the
longitude component of the confined phonon mode has contribution to the
deformation-potential coupling. The piezoelectric coupling is given
by~\cite{PhysRevLett.71.3577}
\begin{equation} 
  H^{\rm{P}}_{\rm{ep}}(\bm{r}) = \frac{e}{\kappa} \int d\bm{r_{\rm{e}}} \frac{ \nabla \cdot
    \bm{P}^{\rm{PZ}}}{|\bm{r} - \bm{r}_{\rm{e}}|},
  \label{eq:pie}
\end{equation}
where $\bm{P}^{\rm{PZ}}$ is the polarization induced by the displacement field and
$\kappa$ is the dielectric constant. For InAs [111] nanowires with wurtzite
structure, $\bm{P}^{\rm{PZ}}$ can be expressed as~\cite{weber}
\begin{eqnarray}
    &&\hspace{-1.2cm} \bm{P}^{\rm{PZ}}(\bm{r}) = e_{15} \Big( \partial_r u_z(\bm{r}) + \partial_z u_r(\bm{r}) \Big) \bm{e}_r  \nonumber\\
    &&\mbox{} + \Big[ e_{31} (\partial_r u_r(\bm{r}) + u_r(\bm{r})/r) + e_{33} \partial_z u_z(\bm{r}) \Big] \bm{e}_z,  
  \label{eq:pz}
\end{eqnarray}
with $e_{15}$, $e_{31}$ and $e_{33}$ being the piezoelectric constants. Note
that due to the anisotropy of the polarization, the piezoelectric coupling can
be more sensitive to the confined phonon mode than the deformation-potential
coupling.

In the following discussion, we restrict to the dilatation modes [which have
angular momentum quantum number $l=0$ and angular component
$\bar{u}^{\theta}_{\nu q}(r)=0$ in Eq.~(\ref{eq:uph})] since only these modes
can couple to the electrons in the QDs in the lowest subband of radial
direction.~\cite{prl.104.036801, PhysRevB.54.1494, comment_dilatation}

For two-electron QDs, the total Hamiltonian of the system is given by
\begin{equation}
  H = (H^{1}_{\rm{e}} + H^{2}_{\rm{e}} + H_{\rm{C}}) + H^{1}_{\rm{ep}} + H^{2}_{\rm{ep}} + H_{\rm{p}},
  \label{eq:h_all}
\end{equation}
where $H_{\rm{C}} = \frac{e^2}{\kappa} \frac{1}{\left| \bm{r}_1 - \bm{r}_2 \right|}$
represents the Coulomb interaction between the two electrons. $H_{\rm{p}} = \sum_{\nu
  q} \hbar \omega_{\nu q} a^{\dagger}_{\nu q } a_{\nu q}$ is the Hamiltonian for
the confined phonons and $H_{\rm{ep}} = H^{\rm{D}}_{\rm{ep}} + H^{\rm{P}}_{\rm{ep}}$ is the electron-phonon
coupling. The superscript ``1'' and ``2'' label the two electrons.

To construct the two-electron basis functions, it is convenient to use the
separation of variables in terms of the center of mass $\bm{R} = (\bm{r}_1 +
\bm{r}_2)/2$ and the relative motion $\bm{r} = \bm{r}_1 - \bm{r}_2$. As the
Coulomb interaction is too strong, we choose the basis function as the
eigenstates of the Hamiltonian $(H^1_{\rm{e}} + H^2_{\rm{e}} + H_{\rm{C}})$ without SOC and Zeeman
term, which can be solved numerically by finite difference method in real
space. The basis functions have the form
\begin{equation}
  \langle \bm{R}, \bm{r} | n_{\rm{R}} n_{\rm{r}} \eta \rangle = \mathcal{R}^{\rm{R}}(R) \mathcal{R}^{\rm{r}}(r)
  \phi^{\rm{R}}_{n_{\rm{R}}}(Z) \phi^{\rm{r}}_{n_{\rm{r}}}(z)
  \mathcal{\chi}_{\eta}, 
  \label{eq:basis}
\end{equation}
where $\mathcal{\chi}_{\eta}$ represents the spin states of the two electrons,
which can be expressed as
\begin{equation}
  \mathcal{\chi}_{\eta} = \left\{ \begin{array}{ll}
      | S \rangle = \frac{1}{\sqrt{2}} ( | \uparrow \downarrow \rangle - | \uparrow
      \downarrow \rangle ), & \eta = 0 \\
      |T_{-} \rangle = | \downarrow \downarrow \rangle, & \eta = 1 \\
      |T_{0} \rangle = \frac{1}{\sqrt{2}} ( | \uparrow \downarrow \rangle + | \uparrow
      \downarrow \rangle ), & \eta = 2 \\
      |T_{+} \rangle = | \uparrow \uparrow \rangle, & \eta = 3 \\
    \end{array} \right..
\end{equation}
In Eq.~(\ref{eq:basis}), $\mathcal{R}^{\rm{R}}(R)$/$\mathcal{R}^{\rm{r}}(r)$ is
the wave function for the lowest subband in radial direction and
$\phi^{\rm{R}}_{n_{\rm{R}}}(Z)$/$\phi^{\rm{r}}_{n_{\rm{r}}}(z)$ is the wave
function in axial direction. The subscript ``$R$'' and ``$r$'' label the center
of mass and relative motion respectively. To guarantee the anti-symmetric of the
wave function, the singlet state $| S \rangle$ always corresponds to
$\phi^{\rm{r}}_{n_{\rm{r}}}(z)$ with even parity, while the triplet states $|
T_{\pm, 0} \rangle$ correspond to $\phi^{\rm{r}}_{n_{\rm{r}}}(z)$ with odd
parity. The energy levels and eigenstates of the two electrons can be obtained
by diagonalizing the full two-electron Hamiltonian $(H^1_{\rm{e}} + H^2_{\rm{e}}
+ H_{\rm{C}})$ in these basis. We identify an eigenstate as singlet and/or
triplet by its expectation value $\langle (\bm{\sigma}_1 + \bm{\sigma}_2)^2
\rangle$.

Treating $| i \rangle$ and $| f \rangle$ as the initial and final states, we can
calculate the phonon-induced relaxation rate using Fermi's golden rule. At zero
temperature, the relaxation rate induced by the confined phonons
reads~\cite{yin}
\begin{equation}
  \Gamma_{fi} = \sum_{j m \nu} \left.\frac{\left| M^j_{\nu q_m}
        \langle f \left| V^j_{\nu q_m} \right| i \rangle  \right|^2}
    {\left|\partial_{q_m} (\hbar \omega_{\nu}(q_m)) \right|}\theta(E)
  \right|_{\hbar \omega_{\nu}(q_m) = E},
  \label{eq:srt}
\end{equation}
where $E = \left| E_f - E_i \right|$ is the energy splitting and $\theta(E)$ is
the step function. $V^j_{\nu q} = W^j_{\nu q}(r_1) e^{i q z_1} + W^j_{\nu
  q}(r_2) e^{i q z_2}$ comes from the total electron-phonon interaction
Hamiltonian $H^1_{\rm{ep}} + H^2_{\rm{ep}}$, with $j=D$ for the
deformation-potential coupling and $j=P$ for the piezoelectric coupling. The
corresponding coefficients are $| M^{\rm{D}}_{\nu q} |^2 = \hbar \Xi^2/( 2 \pi
\rho \omega_{\nu q} R^2 )$ and $| M^{\rm{P}}_{\nu q} |^2 = 16 \hbar \pi^2 e^2
e^2_{14}/( \kappa^2 2 \pi R^2 \omega_{\nu q} )$. The quantity $W^{\rm{D/P}}_{\nu
  q}(r)$ are given in detail in Appendix. Note that the relaxation rate induced
by bulk phonons can be expressed with similar equation as Eq.~(\ref{eq:srt}),
the main difference is that since the constant-energy surface is continuous for
bulk phonons, the discrete summation over $q_m$ should be replaced by the
integration over wave vector $\bm{q}$.

Before we represent the results, we would like to briefly summarize the three
mechanisms which is crucial for the ST relaxation. The first one is the ST
mixing. The ST relaxation favors strong ST mixing where the SOC couples the
states more efficiently. The second one is the ratio between the phonon emission
wave length and the QD size. The phonon emission efficiency is maximized as the
two length is comparable to each other, leading to large relaxation rate. The
third one is the phonon density of states (DOS) at the phonon emission energy,
which is equal to the ST energy splitting. Larger DOS implies that more phonons
can be emitted, which enhances the ST relaxation. The first two mechanisms
manifest themselves in $\langle f \left| V^j_{\nu q_m} \right| i \rangle$ in
Eq.~(\ref{eq:srt}), which is known as the form factor. The third one, e.g., the
phonon DOS, is closely related to the quantity $\partial_{q_m} (\hbar
\omega_{\nu}(q_m))$, which is very sensitive to the phonon spectrum. Thus this
one is expected to lead to pronounced differences in the ST relaxation induced
by confined phonons and bulk phonons.

\section{Results}

In the numerical calculation, we use the parameters for the InAs [111] wurtzite
nanowires.~\cite{fasth, pfund, pfund2} The deformation-potential strength $\Xi$
is chosen to be $5.8$~eV, the density is $\rho=5900$~kg/m$^3$, the static
dielectric constant $\kappa$ is $15.15$, the longitude and transverse sound
velocities are chosen to be $v_L = 4410$~m/s and $v_T = 2130$~m/s
respectively.~\cite{landolt_book} Since the piezoelectric constant for wurtzite
InAs nanowire is largely missing, we use the constants calculated by
transformation from zinc-blende value $e_{14}$, which are $e_{15} = e_{31} = -
e_{33}/2 = -e_{14}/\sqrt{3}$, with $e_{14} = 3.5 \times
10^8$~V$/$m.~\cite{prl.104.036801, bykhovski} The $g$ factor is set to
$-9.0$,~\cite{fasth} and the Rashba coupling constant $\gamma$ is chosen to be
$1.5 \times 10^{-11}$~eV$\cdot$m (corresponding to the SO length
$\lambda_{\rm{SO}} = 200~$nm).~\cite{hanson3} We set the radius of the nanowire
$R$ to $15$~nm. The effective QD diameter $d_0$ is set to $8$~nm. The dot length
$d_z$ is set to $70$~nm.

\begin{figure}
  \includegraphics[width=7.5cm]{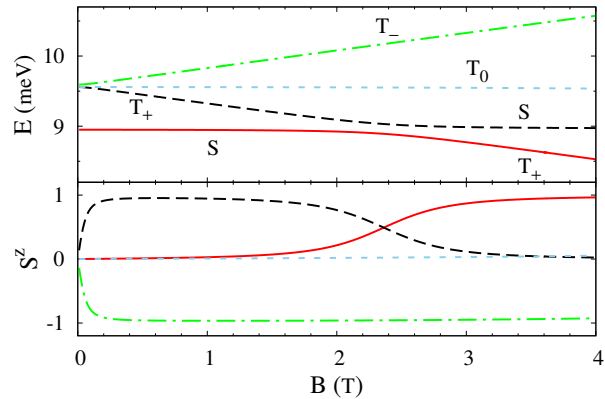}
  \caption{(Color online) The lowest 4 energy levels and expectation value of $S^z=(\sigma^z_1
    + \sigma^z_1)/2$ $v.s.$ external magnetic field $B$.}
  \label{fig:eb}
\end{figure}

We employ the exact diagonalization method with the lowest 48 basis functions to
converge the energy levels and ST relaxation rates.~\cite{jlcheng, jhjiang} The
magnetic field dependence of the first four levels and the corresponding
expectation value of $S^z=(\sigma^z_1 + \sigma^z_1)/2$ are plotted in
Fig.~\ref{fig:eb}(a) and (b) respectively. An anti-crossing at $B=2.36$~T can be
identified in the energy levels. The expectation values $\langle S^z \rangle$
show an crossing at the corresponding magnetic field, indicating a strong mixing
between the singlet and spin up triplet state (ST mixing). As the orbital effect
of the magnetic field is neglected, the levels show linear field dependence,
which is different from the typical disk-shaped QDs.~\cite{fasth, romanon} We
emphasis that the mixing between the singlet and triplet states in our elongate
QDs is more stronger than the QDs studied before.~\cite{climente, shen,
  golovach} This can be seen in the large level splitting, which is $0.14$~meV
at the anti-crossing. This value is much larger than the typical value for
disk-shaped GaAs QDs, which is usually a few $\mu$eV. The strong mixing is not
only due to the strong SOC in InAs, but also due to the large Coulomb
interaction in the elongate QDs, which can enhance the relative strength of
SOC.~\cite{fasth, golovach}

\begin{figure}
  \includegraphics[width=7.5cm]{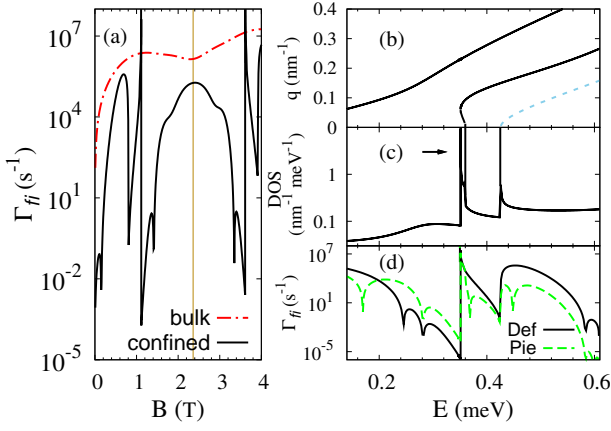}
  \caption{(Color online) (a) ST$_{+}$ relaxation rate as a function of external
    magnetic field. The black solid/red dash-dotted curve represents the
    relaxation induced by confined/bulk phonons. The thin yellow line indicates
    the position of the anti-crossing. (b) The energy spectrum of the relevant
    dilatation modes. Black solid/Blue dotted curves represent the axial/radial
    mode as $q \to 0$. (c) The corresponding phonon DOS. The arrow indicates the
    van Hove singularity with phonon wave vector $q \ne 0$. (d) ST$_{+}$
    relaxation rate induced by confined phonons as a function of energy
    splitting corresponding to the magnetic field region $B \in [0.0, 2.36]$~T
    (before the anti-crossing). Black solid/Green dashed curve represents the
    relaxation due to the deformation-potential/piezoelectric coupling.}
  \label{fig:srt_st+}
\end{figure}

We start our discussion with the ST$_{+}$ relaxation. The magnetic field
dependence of the ST$_{+}$ relaxation is shown in Fig.~\ref{fig:srt_st+}(a). It
can be seen that the relaxation rates induced by bulk and confined phonons are
very different from each other. For bulk phonons, the relaxation rate varies
smoothly as a function of magnetic field. A local minimum can be found at the
anti-crossing. For the confined phonons, the relaxation rate exhibits
multi-peaks as a function of magnetic field. The two ``sharp'' peaks located at
$B=1.11$~T and $B=3.62$~T are actually two divergent peaks. Moreover, the
relaxation rate exhibits a local maximum at the anti-crossing in contrast to the
local minimum for the bulk phonons. We attribute the multi-peak structure and
the local maximum at the anti-crossing as the two main features of the ST$_{+}$
relaxation rate induced by confined phonons.

The first feature, e.g., the multi-peak structure of the relaxation rate, is a
direct consequence of the large DOS at the van Hove singularities in the
confined phonon DOS. To see this, we plot the spectrum of the relevant confined
phonon modes and the corresponding DOS in Fig.~\ref{fig:srt_st+}(b) and (c)
respectively. The ST$_{+}$ relaxation induced by the confined phonons via the
deformation-potential coupling and the piezoelectric coupling are plotted as a
function of energy splitting in Fig.~\ref{fig:srt_st+}(d) [Here we only plot the
relaxation rate before the anti-crossing for clarification, the relaxation after
the anti-crossing has similar behavior]. It is easy to see that the peaks in the
relaxation rate correspond to the van Hove singularities in the DOS, indicating
the strong enhancement due to the large DOS. The divergent peaks correspond to
the van Hove singularity with $q \ne 0$, while the non-divergent peaks are
induced by the van Hove singularity with $q = 0$. This is because as $q \to 0$,
the form factor tends to zero, which will suppress the relaxation
rate.~\cite{yin} Note that the form factor can also have zeros for $q \ne 0$,
which results in the dips in the relaxation rate. There are two types of
zeros. One is the zeros due to the axial component (corresponding to common
peaks for both the deformation-potential coupling and the piezoelectric
coupling),~\cite{bulaev} the other one is the zeros due to the radial component,
which induces different dips for different electron-phonon coupling
mechanism. Note that the deformation-potential coupling dominates the relaxation
rate in most cases except in the energy region $[0.2, 0.35]$~meV. This is
because in this region, the corresponding phonon mode tends to be transverse
mode which decouples to the electrons via the deformation-potential
coupling.~\cite{yin}

It can be seen for Fig.~\ref{fig:srt_st+} that the second feature, e.g., the
local maximum of the relaxation rate, occurs in the region far away from the van
Hove singularities, where the confined phonon DOS exhibits a plateau. In this
region, the behavior of the relaxation rate is decided by the combined effect of
the ST mixing and the phonon emission efficiency. Both mechanisms enhance the
relaxation rate at the anti-crossing, resulting in the local maximum. For bulk
phonons, the DOS is a quadratic function of the energy splitting, so the
relaxation rate can be suppressed as energy splitting decreasing, resulting in
the local minimum at the anti-crossing. Note that the energy splitting $E$ here
($E = 0.14$~meV) is much larger than the energy splitting in typical disk-shaped
GaAs QDs, so the suppression of the relaxation rate due to the decreasing of the
phonon emission efficiency is absent.

To further justify the behavior of the relaxation rate at the anti-crossing, we
give an estimation based on perturbation theory. We concentrate on the
deformation-potential coupling since it dominates the ST relaxation rate at the
vicinity of the anti-crossing. We restrict the discussion with the lowest two
basis $|0 0 S\rangle$ and $|0 1 T_{+}\rangle$ with energy $\epsilon_0$ and
$\epsilon_1 + g \mu_{\rm{B}} B$ since they are almost degenerated at the
vicinity of the anti-crossing. By applying the degenerated perturbation theory,
we get the eigenstates
\begin{eqnarray}
  | \bar{S} \rangle & = & \cos\theta | 00 S \rangle + \sin\theta | 01 T_{+} \rangle, \\
  | \bar{T}_{+} \rangle & = & - \sin\theta | 00 S \rangle + \cos\theta | 01 T_{+} \rangle,
  \label{eq:eig_st}
\end{eqnarray}
where $\tan\theta = t/(d + \sqrt{d^2 + t^2})$ with $t= \gamma \alpha_z\langle
\phi^r_0(z) | \partial_z | \phi^r_1(z) \rangle$ and $d=(\epsilon_0 - \epsilon_1
+ g \mu_{\rm{B}} B)/2$.  The energy splitting between the two eigenstates is
$E=2 \sqrt{d^2 + t^2}$. Note that the two states are maximally mixed as $E$
reaches its minimum.

The relaxation between the singlet state $| \bar{S} \rangle$ and triplet state
$| \bar{T}_{+} \rangle$ can be calculated by applying Eq.~\ref{eq:srt}. For bulk
phonons, we have
\begin{eqnarray}
    \Gamma^{\rm{bulk}}_{\rm{ST_{+}}} & = &E \left| M^{\rm{D}}_{\rm{bulk}} \right|^2 t^2
    \int^\pi_0 \sin\theta d\theta \left| I^z_{\rm{ST_{+}}}(q_c \cos\theta) \right|^2  \nonumber\\
    &&\mbox{} \times e^{-(5/4 \pi)(q_c d_0 \sin\theta)^2},
  \label{eq:srt_bulk}
\end{eqnarray}
where $\left| M^{\rm{D}}_{\rm{bulk}} \right|^2 = \sqrt{\pi} D^2/(4 d_z^3 \hbar^4
v_L^5 \rho)$ is a constant. $q_c = E/(\hbar v_L)$ is the bulk phonon emission
wave vector. $I^z_{\rm{ST_{+}}}(q) = \langle 00S | e^{i q z_1} + e^{i q z_2}
|00S \rangle - \langle 01T_{+} | e^{i q z_1} + e^{i q z_2} | 01T_{+} \rangle$ is
the axial component of the form factor. In the situation we considered here,
$I^z_{\rm{ST_{+}}}(q)$ varies slowly, so we approximate it as a constant. For
small $E$, one has
\begin{equation}
  \Gamma^{\rm{bulk}}_{\rm{ST_{+}}} \propto E,
  \label{eq:srt_bulk2} 
\end{equation}

For confined phonons, only the lowest confined phonon mode has contribution at
the vicinity of the anti-crossing, so we have
\begin{eqnarray}
  &&\hspace{-1.2cm} \Gamma^{\rm{conf}}_{\rm{ST_{+}}} = E^{-1} \left| M^{\rm{D}}_{\rm{conf}}
  \right|^2 t^2 \left| I^z_{\rm{ST_{+}}}(q_c
    ) \right|^2 \left. \frac{\partial q}{\partial \omega_{1 q}} \right|_{q=q_c} \nonumber\\
  &&\mbox{} \times \left| I^{d}_{\rm{conf}} (q_c, \sqrt{2} d_0) I^{d}_{\rm{conf}} (q_c,
    d_0/\sqrt{2}) \right|^2,
  \label{eq:srt_conf}
\end{eqnarray}
where $\left| M^{\rm{D}}_{\rm{conf}} \right|^2 = D^2 \omega^4_z/ (2 \sqrt{\pi}
d_z \rho R )$ is a constant. Note that the eigenenergy of the first confined
phonon mode tends to be linear in $q$, so we can write $q_c = E/(\hbar
v_S)$.~\cite{auld} $I^z_{\rm{ST_{+}}}(q)$ is the axial form factor which is the
same as the bulk phonons since both the bulk and confined phonons can be
expressed as plain wave in axial direction. $I^d_{\rm{conf}}(q, d_0) =
\chi^{(0)}_{\nu q} ( k_L / d_0 )^2 e^{A_L(\nu, q)}/\sqrt{\omega_{\nu
    q}/\omega_z}$ is the radial form factor for confined phonons, with $k_L =
\omega_{\nu q}/v_L$ and $A_L(\nu, q) = {(q^2-k^2_L)
  d_0^2}/{(4\pi)}$. $\chi^{(0)}_{\nu q}$ is the coefficient in the expression
for the confined phonon eigenmode which is calculated numerically. It can be
shown numerically that $I^d_{\rm{conf}}(q, d_0)$ tends to a constant for small
$q$.~\cite{yin} So for small $E$, we have
\begin{equation} 
  \Gamma^{\rm{conf}}_{\rm{ST_{+}}} \propto E^{-1} .
  \label{eq:srt_conf2} 
\end{equation}
Comparing Eq.~(\ref{eq:srt_bulk2}) and Eq.~(\ref{eq:srt_conf2}), one can see
that as the energy splitting $E$ decreasing, the relaxation rate induced by bulk
phonons decreases while the relaxation rate induced by confined phonons
increases.

\begin{figure}
  \includegraphics[width=7.5cm]{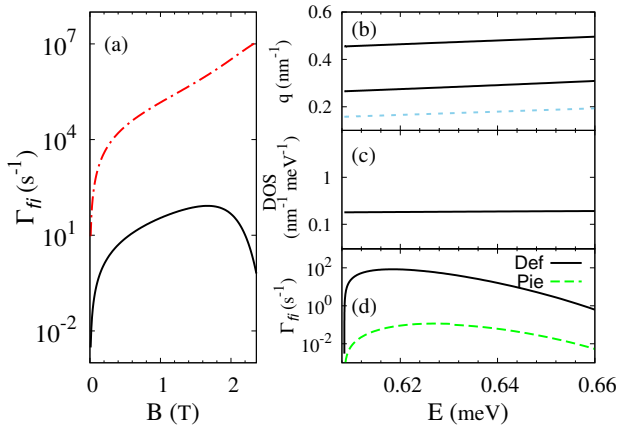}
  \caption{(Color online) Same as Fig.~\ref{fig:srt_st+} but for the ST$_{-}$ relaxation.}
  \label{fig:srt_st-}
\end{figure}

Similar features also exists for ST$_{-}$ and ST$_{0}$ relaxation rates. The
multi-peak structure in the relaxation rate due to van Hove singularities can be
found for ST$_{-}$ relaxation rate shown in Fig.~\ref{fig:srt_st-}. Note that
the ``sharp'' peaks here (the first, second and the fourth peaks) are not
divergent peaks since they correspond to the $q=0$ van Hove singularities.  They
are ``sharp'' since the piezoelectric coupling has a large contribution at
these peaks which can be seen in Fig.~\ref{fig:srt_st-}(d). This is because
these peaks are axial mode as $q \to 0$, which is favored by the piezoelectric
coupling.~\cite{weber} Also note that for bulk phonons, the decreasing of the
phonon emission efficiency due to the small phonon wave length dominates in this
region. So although the DOS increasing as the magnetic field, the relaxation
rate is suppressed.

\begin{figure}
  \includegraphics[width=7.5cm]{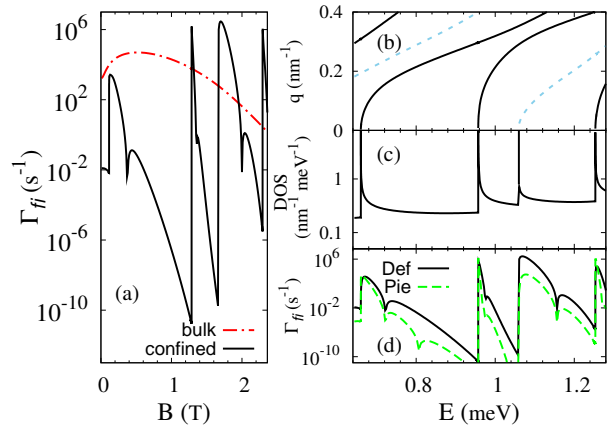}
  \caption{(Color online) Same as Fig.~\ref{fig:srt_st+} but for the ST$_{0}$ relaxation.}
  \label{fig:srt_st0}
\end{figure}

In the region where the confined phonon DOS exhibits plateaus, the effect of
phonon emission efficiency can be pronounced, which may suppress the relaxation
rate. As an example, we compare the ST$_{0}$ relaxation rate induced by bulk and
confined phonons in Fig.~\ref{fig:srt_st0}. For large magnetic field, the
relaxation rate is suppressed due to the decreasing of the phonon emission
efficiency for confined phonons. For bulk phonons, the relaxation rate is still
increasing due to the increasing of the phonon DOS. Note that the increasing of
the relaxation rate for both confined phonons and bulk phonons in small magnetic
field is due to the ST mixing.

\section{Summary}

In summary, we have investigate the ST relaxation induced by the confined
phonons in two-electron nanowire-based QDs. We find that the behavior of the
relaxation rate is dominated by the large DOS at the vicinity of the van Hove
singularities, while in the region where the confined phonon DOS is flat, the
effect of the phonon emission efficiency and the ST mixing are more
pronounced. This results in the multi-peak structure of the ST relaxation rate
and the local maximum of ST$_{+}$ relaxation rate at the vicinity of the
ST$_{+}$ anti-crossing. These features are very different from the previous
results for disk-shaped QDs, indicating the unique property of the confined
phonons. The effect of the deformation-potential coupling and piezoelectric
coupling are also discussed. We find that for InAs [111] nanowire, the
deformation-potential coupling dominates the relaxation in most cases except in
the region where the longitude component of the confined phonons mode is
suppressed. The piezoelectric coupling is also found to be important for axial
confined phonon mode at $q \to 0$. It is also worth noting that the relaxation
rate induced by confined phonons is much smaller than the one induced by bulk
phonons in most cases, this suggests the nanowire-based QDs are preferable for
the application in quantum information and computation.

\begin{acknowledgments}
 The author would like to thank M. W. Wu for proposing the topic as
  well as the directions during the whole investigation.
This work was supported by the Natural Science Foundation of China under Grant
No.~10725417, the National Basic Research Program of China under Grant
No.~2006CB922005 and the Knowledge Innovation Project of Chinese Academy of
Sciences and also partially by the China Postdoctoral Science
Foundation.
\end{acknowledgments}

\begin{appendix}
  \section{$W^{\rm{D/P}}_{\nu q} (r)$ in Eq.~\ref{eq:srt}}
  The quantity $W^{\rm{D/P}}_{\nu q} (r)$ is the radial component of the
  electron-phonon coupling $H^{\rm{D}/\rm{P}}_{\rm{ep}}$ with the superscript
  ``D'' for the deformation-potential coupling and ``P'' for the piezoelectric
  coupling. They can be obtained by substituting Eq.~(\ref{eq:uph}) into
  Eq.~(\ref{eq:def})/Eq.~(\ref{eq:pie}).
  
  For the deformation-potential coupling, $W^{\rm{D}}_{\nu q} (r)$ is given by
  \begin{equation}
    W^{\rm{D}}_{\nu q} (r) = ( k^2_L + q^2 ) J_0 (k_L r) \chi^{(0)}_{\nu q},
  \end{equation}
  where $k^2_{L/T} = (\omega_{\nu q}/v_{L/T})^2 - q^2$. 
  
  For the piezoelectric coupling for wurtzite InAs [111] nanowire,
  $W^{\rm{P}}_{\nu q} (r)$ is given by
  \begin{eqnarray}
      &&\hspace{-1.4cm} W^{\rm{P}}_{\nu q} (r) = \int r' dr' I_0(q r_{<}) K_0(q r_{>}) \nonumber\\
      &&\mbox{} \times q \Big[ \chi^{(0)}_{\nu q} J_0 (k_L r') (2
      q^2 - 3 k^2_L) \nonumber\\
      &&\mbox{} + \chi^{(2)}_{\nu q} J_0 (k_T r') (k^2_T -4q^2)/q^2 \Big]/\sqrt{3}.
  \end{eqnarray}
  where $r_{<} = \min(r,r')$ and $r_{>} =
  \max(r,r')$. $J_0(r)$($I_0(r)$/$K_0(r)$) is the zeroth-order (Modified) Bessel
  function. $\chi^{(0)}_{\nu q}$ and $\chi^{(2)}_{\nu q}$ are coefficients in
  the expression for the confined phonon eigenmode which are calculated
  numerically following Refs.~\onlinecite{auld,stroscio}.
  
\end{appendix}


\begin{thebibliography}{46}
\bibitem{comment_dilatation} Note that this is due to the fact that the InAs
  [111] nanowire is wurtzite structure. For zinc-blende nanowire, piezoelectric
  coupling can also couple to phonon modes with $l=\pm2$.
\end{thebibliography}
\end{document}